\documentclass[reprint,11pt]{revtex4}
\usepackage{xfrac}

\begin{document}
\title{Bose-Einstein Condensation of Photons versus Lasing and Hanbury Brown-Twiss Measurements with a Condensate of Light}

\author{Julian Schmitt, Tobias Damm, David Dung, Frank Vewinger, Jan Klaers\footnote{Present address: Institute for Quantum Electronics, ETH Zurich, Auguste-Piccard-Hof 1, 8093 Zurich, Switzerland}, and Martin Weitz$^\dagger$}

\address{Institut f\"ur Angewandte Physik, Universit\"at Bonn,\\
Wegelerstr. 8, 53115 Bonn, Germany\\
$^\dagger$E-mail: martin.weitz@uni-bonn.de}

\begin{abstract}
The advent of controlled experimental accessibility of Bose-Einstein condensates, as realized with e.g. cold atomic gases, exciton-polaritons, and more recently photons in a dye-filled optical microcavity, has paved the way for new studies and tests of a plethora of fundamental concepts in quantum physics. We here describe recent experiments studying a transition between laser-like dynamics and Bose-Einstein condensation of photons in the dye microcavity system. Further, measurements of the second-order coherence of the photon condensate are presented. In the condensed state we observe photon number fluctuations of order of the total particle number, as understood from effective particle exchange with the photo-excitable dye molecules. The observed intensity fluctuation properties give evidence for Bose-Einstein condensation occurring in the grand-canonical statistical ensemble regime.
\end{abstract}


\maketitle


\section{Introduction}

For material particles of integer spin (bosons), Bose-Einstein condensation to a macroscopically occupied ground state minimizes the free energy when cooled to very low temperatures at sufficient density~\cite{Leggett}. Other than for gases of material particles~\cite{Bongs,Deng,Kasprzak,Balili,Demokritov}, Bose-Einstein condensation usually does not occur for photons~\cite{Huang}. In the most well-known photon gas, blackbody radiation, photons disappear in the system walls when cooled to low temperature instead of exhibiting Bose-Einstein condensation to the ground mode. This is also expressed by the common statement that the chemical potential for photons vanishes. Early theoretical work has proposed Bose-Einstein condensation of photons in the Compton scattering of X-rays~\cite{Zeldovich}, and more recently Chiao proposed a two-dimensional photon fluid in a nonlinear resonator~\cite{Chiao}. Besides in blackbody radiation, thermalization effects have long been accounted for in the description of multimode intracavity spectroscopy laser setups below the laser threshold~\cite{Mironenko}. More recently, Bose-Einstein condensation of exciton-polaritons, which are mixed states of matter and light in the strongly bound limit, has been experimentally achieved~\cite{Deng,Kasprzak,Balili}. Here the material part of the polaritons drive the system into or near thermal equilibrium. Our group in 2010 observed Bose-Einstein condensation of photons in a dye-filled optical microcavity~\cite{Klaers1}, see also recent work by Marelic and Nyman~\cite{Marelic}. A very short optical cavity here imprints an effective low-frequency cutoff for photons, with a spectrum of allowed photon energies well above the thermal energy in frequency units. Thermalization of the photon gas to the (rovibrational) temperature of the gas proceeds by absorption and re-emission processes on the dye molecules. For corresponding theoretical works, see~\cite{Leeuw,Snoke,Kirton,Kruchkov,Zhang,Klaers2,Sobyanin,Strinati}.\\

This article reviews recent experiments of our group studying the transition from laser-like dynamics to Bose-Einstein condensation of photons upon variation of the thermalization rates of the photon gas to the dye medium~\cite{Schmitt1}. Moreover, we describe work observing grand-canonical number statistics of the condensate emission, as understood from effective particle exchange with the reservoir of photo-excitable dye molecules~\cite{Klaers2,Sobyanin,Schmitt2}.\\

In the following, Chapter 2 describes the experimental dye microcavity system, in which the Bose-Einstein condensate is generated, and Chapter 3 presents experiments studying a crossover between laser-like nonequilibrium dynamics and Bose-Einstein condensation of photons. Further, Chapter 4 gives results obtained from studying the condensate number statistics and Chapter 5 concludes this article.\\

\section{Two-Dimensional Photon Gas in Microcavity}

Our experimental approach utilizes a thermal coupling of a two-dimensional photon gas to a bath of dye molecules. A simplified schematic of the experiment, which has been previously discussed in detail e.g. in Ref.~\cite{Klaers5}, is shown in Fig.~1(a). The experiments are conducted in a cavity consisting of two curved mirrors spaced in the micrometer regime, filled with dye in liquid solution. The mirrors, due to their small spacing, impose an upper limit to the optical wavelength that fits into the cavity, corresponding to a restriction of energies to a minimum cutoff of $\hbar\omega_\textrm{c}\approx 2.1~\textrm{eV}$, which is much larger than thermal energy $k_\textrm{\tiny B}T \approx 1/40~\textrm{eV}$ at room temperature ($T=300~\textrm{K}$). In this case, thermal emission of photons into the modes of the cavity is suppressed by a factor of order $\exp(-\hbar\omega_\textrm{c}/k_\textrm{B}T) \approx 10^{-36}$ with the above given numbers, which is a precondition for an independent tuning of photon number and temperature. Note that the limit $\hbar\omega\gg k_\textrm{B} T$ is as well fulfilled in usual laser physics, while this separation of energy scales is not fulfilled for a blackbody radiator. By repeated absorption re-emission processes, the photons thermalize to the rovibrational temperature of the dye, which corresponds to room temperature. In the course of thermalization, the longitudinal modal quantum number of the photons in the thin cavity remains fixed, and the photon dynamics is restricted to the remaining two modal degrees of freedom. In thermal equilibrium, the photon frequencies will then be distributed by ${\simeq}~k_\textrm{\tiny B} T/\hbar$ above the low-frequency cutoff. Rapid decoherence from collisions with solvent molecules prevents a coupling of the phases of the dye molecular dipole and the photon, so that we can well assume that in our experiments photons instead of polaritons are studied.\\

\begin{figure}[t]
\begin{centering}
\vspace{0mm}
\includegraphics[width=1.0\textwidth]{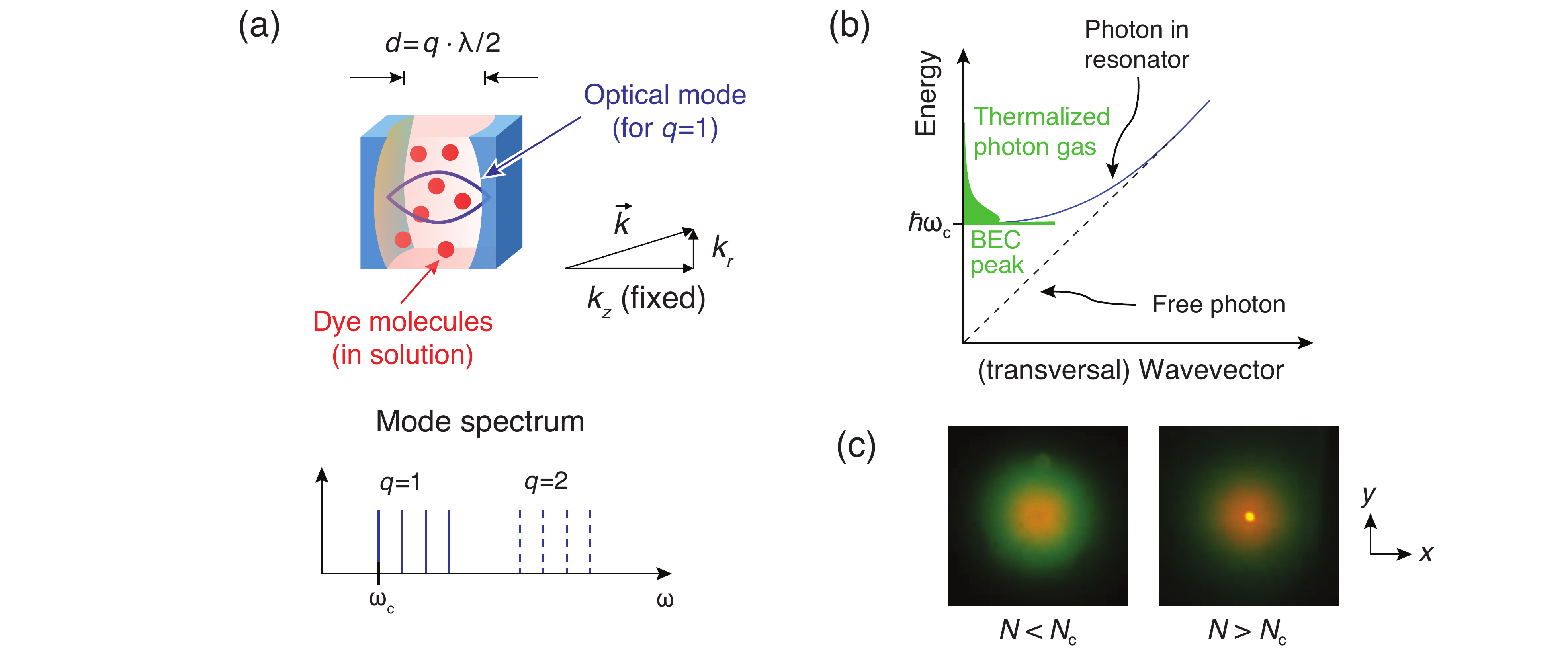}
\par\end{centering}
\vspace{0mm}
\caption{(a) Scheme of optical microresonator (top) and mode spectrum (bottom) for the case of a spacing between resonator mirrors of half an optical wavelength ($q=1$). The resonator is filled with dye solution and the photons are thermally coupled to the dye by repeated absorption-emission processes. (b) Photon dispersion in resonator (solid line) and the dispersion of a free photon (dashed line). (c) Spatial images of the cavity emission both in the thermal (left) and in the condensed regime (right), the latter with the bright spot in the center corresponding to the BEC peak.}
\vspace{0mm}
\end{figure}

Inside the resonator the optical dispersion becomes quadratic, see Fig.~1(b), and the photon gas behaves equivalent to a two-dimensional gas of massive bosons that is harmonically confined, with the latter being caused by the curvature of the cavity mirrors. In contrast to a homogeneous two-dimensional Bose gas, Bose-Einstein condensation here is possible~\cite{Bagnato}. Interestingly, the effective mass $m_\textrm{ph} = \hbar\omega_\textrm{c}/c^2$, where $c$ denotes the speed of light in the medium and $\omega_\textrm{c}$ the cutoff frequency, is some 10 orders of magnitude smaller than the mass of alkali atoms, and the Bose-Einstein condensation transition temperature can be at room temperature. In our experiment, an initial photon population is injected into the dye cavity system by pumping the dye with an external laser beam, either in a temporally pulsed or quasi-cw way. In the latter case, losses from the 'photon box' can be compensated for by maintaining the molecular excitation level of the dye at a constant level. We have experimentally observed both the thermalization~\cite{Klaers3}, as well as Bose-Einstein condensation of photons~\cite{Klaers1}. Figure~1(c) shows typical images for the emitted radiation transmitted through a cavity mirror below (left) and above the phase transition to a Bose-Einstein condensate (right). We observe photon gases with typically up to 70\% condensate fraction very closely following expectations for a thermal equilibrium distribution. Evidence for a BEC of photons was obtained from the observed spectra showing Bose-Einstein distributed photon energies with a macroscopically occupied peak on top of a thermal cloud, the observed threshold of the phase transition showing the predicted absolute value and scaling with e.g. mirror curvature, and condensation in the trap center even for a spatially offset pump beam, as possible by the thermalization~\cite{Klaers1,Schmitt1}.\\

\section{Nonequilibrium Lasing versus Equilibrium Condensation of Photons}

Trapped dilute cold atomic gas systems achieve a state that is very close to that described by a thermal equilibrium distribution~\cite{Leggett,Bongs}. The establishment of thermal equilibrium conditions demands that there are separate timescales for thermalization with respect to that of losses and pump terms, being a precondition for the concept of Bose-Einstein condensation. In the field of exciton-polaritons, where lifetimes of the quasiparticles typically are as short as several picoseconds, the question whether a system that is pumped and exhibits losses can show Bose-Einstein condensation has been discussed~\cite{Wouters,Deng2}. Experiments give evidence for condensation despite the short lifetime of polaritons~\cite{Deng,Kasprzak,Balili}.\\

As the here investigated system is situated in the regime of weak coupling of matter and light, rate equations can be used to describe the dynamics of photons in the dye microcavity. Usual laser equations, in the limit of negligible loss and assuming that the Kennard-Stepanov law predicting a thermodynamic Boltzmann-type scaling between the Einstein coefficients for absorption and emission respectively holds, yield a thermal distribution of photons in the cavity~\cite{Klaers4}. The intermediate case of particle equilibrium of the photon gas has in detail been theoretically investigated by Kirton and Keeling~\cite{Kirton}. 
Figure~2 gives a comparison of different states of cavity photons, assuming $\hbar\omega\gg k_\textrm{\tiny B} T$. The left half of the diagram refers to a pumped medium with no or negligible thermalization, where below laser threshold the emission is spontaneous, while when inversion is reached at the point when the gain per cavity round trip exceeds the loss laser operation sets in, yielding a macroscopic occupation of modes independent of energetics~\cite{Siegman}. If the cavity lifetime $\tau_\textrm{cav}$ exceeds the characteristic thermalization time of the photon gas $\tau_\textrm{th}$ (right hand side of diagram), a trapped photon will be reabsorbed in the dye heat bath and thermalize before being lost by e.g. mirror transmission. Below the critical photon number for Bose-Einstein condensation, again spontaneous processes dominate but a thermalized distribution of cavity modes is now expected. Once the threshold for a BEC is reached, the lowest energetic mode will be macroscopically populated, forming the BEC peak. Both for the laser and the BEC case Bose-enhancement of modes plays a dominant role in the macroscopic population, a process equally well important in the formation of an atomic condensate, with stimulated scattering processes of atoms in the latter case~\cite{Lee}.\\

\begin{figure}[t]
\begin{centering}
\vspace{0mm}
\includegraphics[width=1.0\textwidth]{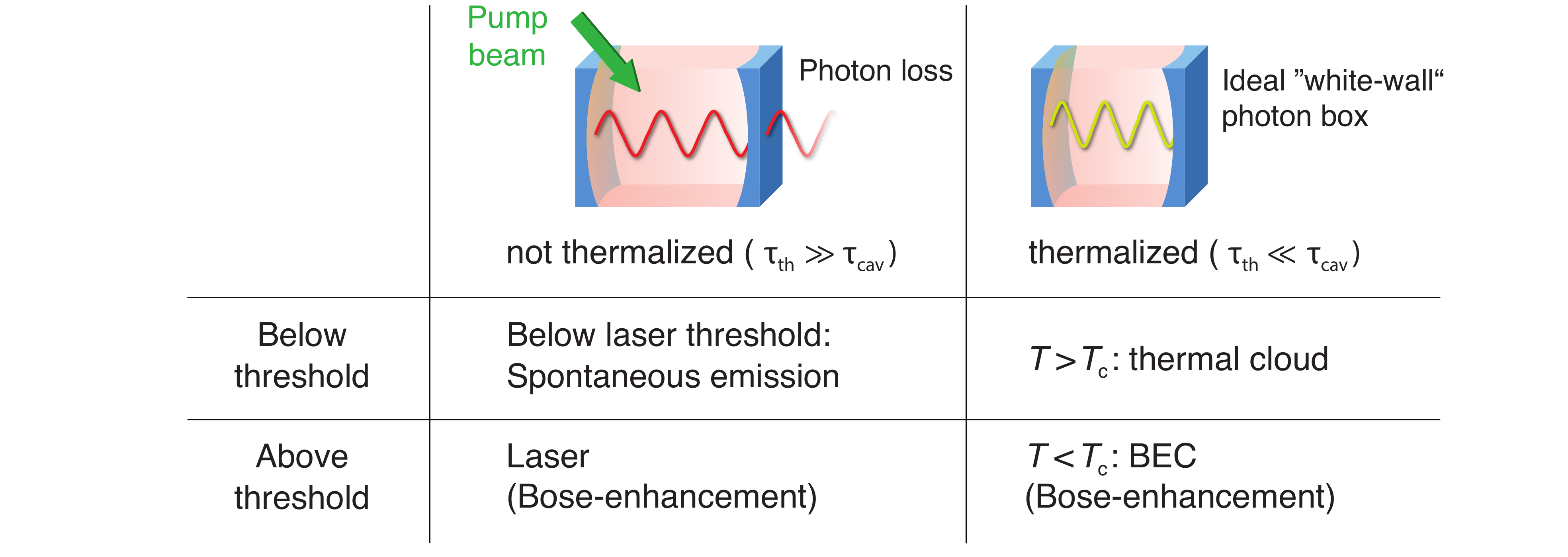}
\par\end{centering}
\vspace{0mm}
\caption{Classification of optical sources far from and at thermal equilibrium respectively, in the regime of the photon energy $\hbar\omega$, being far above the thermal energy $k_\textrm{\tiny B}T$. We assume that the cavity has a low-frequency cutoff and that the thermalization process conserves the average photon number. Thermal equilibrium is obtained also for a pumped system when a photon thermalizes faster than it is lost by e.g. mirror transmission ($\tau_\textrm{th} \ll \tau_\textrm{cav}$, right), while in the opposite limit the state remains far from thermal equilibrium, and becomes laser-like when the gain per cavity roundtrip is higher than photon loss ($\tau_\textrm{th} \gg \tau_\textrm{cav}$, left).}
\vspace{0mm}
\end{figure}

We have experimentally determined the characteristic thermalization time in the dye microcavity system by pumping the dye medium with a pulsed picosecond laser and analyzing the subsequent cavity emission in a time-resolved way using a streak-camera~\cite{Schmitt1}. Under spatially homogeneous excitation of the dye molecules the temporal evolution of the spectral photon distribution reveals the thermalization time after which the spectrum is described by a 300K Bose-Einstein distribution, which is of order of the reabsorption timescale (for typical parameters approximately $20~\textrm{ps}$) of a photon in the dye medium. This indicates that indeed thermalization occurs to the molecular bath. Further, we studied the transition between laser dynamics and equilibrium condensation of photons, for which a focused pump pulse was irradiated spatially removed from the trap center, as to energetically remove the initial state of cavity photons from the low frequency ground state, see Fig.~3 for corresponding data. In the course of our experiments, the cavity cutoff was varied in order to tune the coupling strength to the dye bath, allowing to control the thermalization time of the photons. The data shown on the left hand side of Fig.~3 was recorded for a position of the cavity cutoff $\lambda_\textrm{c} = 596~\textrm{nm}$, for which the wavelength of cavity photons is detuned relatively far from the dye zero-phonon line, so that dye absorption and thermal contact of photons to the dye is weak. We experimentally observe the oscillation of cavity photons within the trapping potential imprinted by the mirror curvature. The photons remain in a non-thermal state at high energies far above the cutoff-frequency, and the according dynamics resembles an analogue to (mode-locked) laser oscillation. Photons here leak out of the cavity before they have a chance to thermalize. The right hand side of Fig.~3 shows data recorded for a cavity cutoff $\lambda_\textrm{c} = 571~\textrm{nm}$, tuning cavity photons in a wavelength range where they are reabsorbed from the traversing wave packet by dye molecules and thermal contact is established. We observe that with advancing times the photon gas accumulates in the trap center and here forms a Bose-Einstein condensate. The observations are consistent with further measurements of the time evolution of the photon spectra. While for weak coupling to the dye heat bath the photon spectrum persists in its initial nonequilibrium state, a spectral redistribution of fluorescence to an equilibrium Bose-Einstein distribution is observed, when thermal contact between photons and dye molecules is enhanced~\cite{Schmitt1}.\\

\begin{figure}[t]
\begin{centering}
\vspace{0mm}
\includegraphics[width=1.0\textwidth]{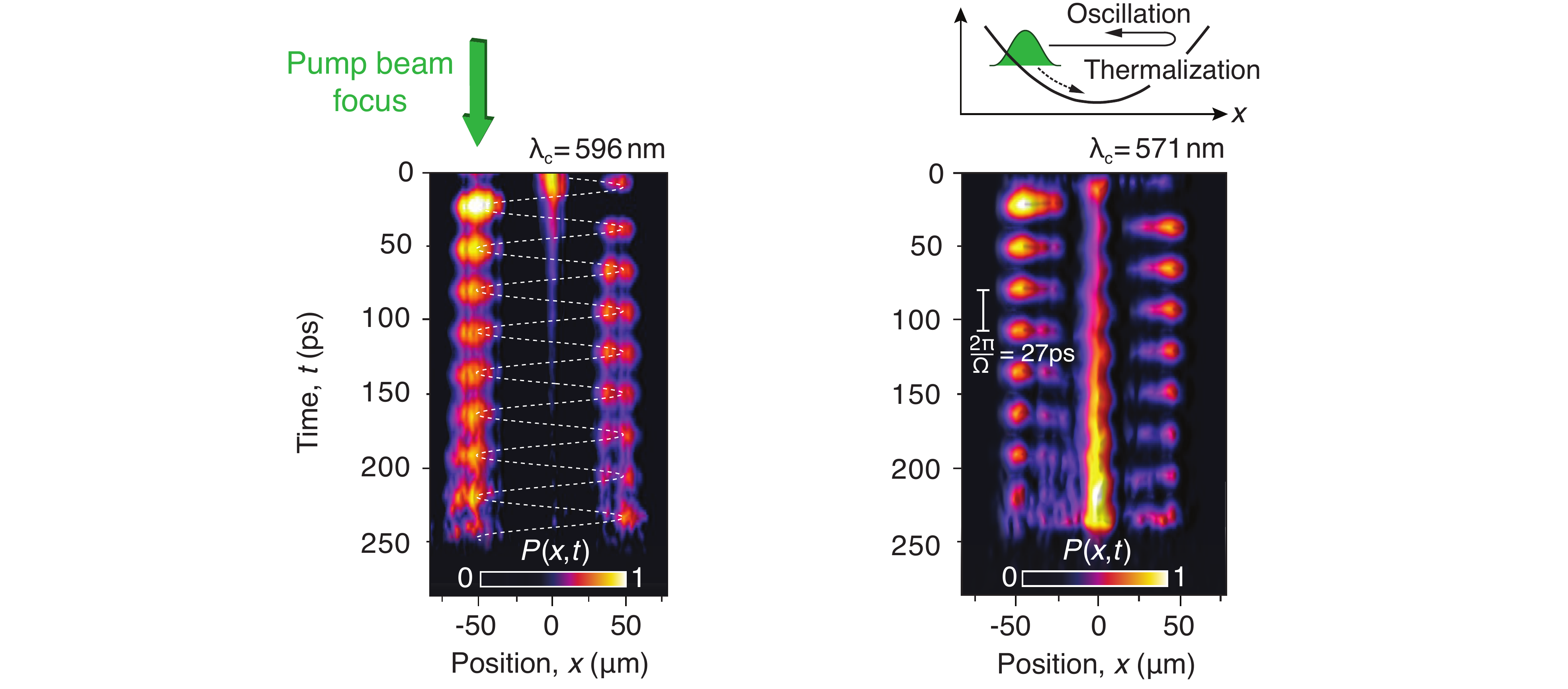}
\par\end{centering}
\vspace{0mm}
\caption{Temporal evolution of the spatial profile (line normalized) of the radiation transmitted through one cavity mirror for two cutoff wavelengths. A pump beam with $27~\mu\textrm{m}$ diameter spatially displaced by $50~\mu\textrm{m}$ from the trap center excites an optical wave packet oscillating in the harmonic trapping potential (indicated on top, right). For the data recorded with shorter cutoff wavelength (right) and correspondingly larger dye absorption a photon condensate in the trap center gradually builds up, while for low reabsorption no condensate emerges (left). }
\vspace{0mm}
\end{figure}

\section{Hanbury~Brown-Twiss~Measurements~of~the~Photon~Condensate}

The above described studies have demonstrated radiative coupling of photons to a thermal heat bath in the composite dye-photon system. Evidently, the photon gas can not be considered as a system isolated from its environment in the sense of the microcanonical statistical ensemble. This leads us to the question, whether the electronically excited dye molecules can additionally constitute an effective reservoir species for the photons, allowing for particle exchange among the two subsystems. In the present section, we describe corresponding experimental work determining the intensity correlations of the dye microcavity emission. The main result is that we observe number fluctuations in the condensed state which are of same order as the average particle number of the condensate. This gives evidence for the condensation occurring in a system described by grand-canonical statistical conditions, due to possible effective particle exchange with between photons and dye electronic excitations.\\

In general, different statistical ensembles represent different conservation laws that can be realized in nature. The microcanonical and canonical ensemble, respectively, refer to physical systems with a fixed number of particles, while energy is fixed in the former case and allowed to fluctuate around a mean value, determined by contact to a heat reservoir, in the latter case. In the grand-canonical ensemble both energy and particle number of a system can vary due to contact with a particle and energy reservoir and one here finds relative fluctuations of all single particle levels of 100\%. For most problems in statistical physics one assumes that the different statistical ensembles become interchangeable in the thermodynamic limit, meaning that relative fluctuations vanish, i.e. $\Delta N/\bar N\rightarrow 0$, with $\bar N$ as the average total particle number and $\Delta N$ its rms fluctuations. Notably, the ideal Bose gas represents an exemption from this generalization, as assuming grand-canonical conditions for the case of the macroscopically occupied ground state present in the Bose-Einstein condensed case yields statistical fluctuations of order of the total particle number, i.e. $\Delta N \approx \bar N$. The fluctuations here do not freeze out at low temperature, instead the prediction is just the opposite: the size of the fluctuations approaches the average particle number as the condensate fraction reaches unity. This counterintuitive phenomenon is commonly referred to as the 'grand-canonical fluctuation catastrophe'~\cite{Fujiwara,Ziff,Holthaus}.\\

Grand-canonical ensemble conditions do not apply to a cloud of an ultracold atomic gases well isolated from the environment, as well as for present polariton condensation experiments~\cite{Deng,Kasprzak,Balili}. The situation however is less obvious for a spatially finite region within a large reservoir, as was first discussed in relation to e.g. liquid helium systems. The physical significance of the grand-canonical ensemble in the condensed phase has long been an open question. Ziff, Uhlenbeck and Kac showed that, for a system in diffusive contact with a spatially separated particle reservoir, the grand-canonical ensemble loses its validity~\cite{Ziff}. These arguments however do not necessarily apply for other types of reservoirs. In the here discussed dye microcavity system, the dye molecules act both as a heat bath and a particle reservoir for the photon gas in a grand-canonical sense, with the contact between system and reservoir not being realized diffusively but by absorption and emission processes, in a spatially overlapping geometry. We have predicted grand-canonical number fluctuations for this system~\cite{Klaers2}, and the obtained theory results for the photon number distribution have been confirmed~\cite{Sobyanin}.\\

To experimentally observe the intensity correlations of the condensate mode of dye microcavity emission, we use a Hanbury Brown-Twiss setup~\cite{Schmitt2}. This experiment is carried out by pumping dye microcavity with typically $150~\textrm{ns}$ long pulses derived by acousto-optically chopping the emission of a cw laser, which is much longer than the picosecond thermalization time~\cite{Schmitt1}. The experiment thus is operated in a 'quasi'-cw mode; as for the case for our initial works described in Section 2. Part of the radiation transmitted through one cavity mirror is spatially filtered in the far-field Fourier plane to separate the condensate mode from the higher transverse modes. Subsequently, the filtered radiation (condensate mode) is split and directed onto two single-photon avalanche photodiodes. A correlation system records time histograms of detection events at the detectors, from which intensity correlations $g^{(2)}(\tau)$ are determined.\\

\begin{figure}[t]
\begin{centering}
\vspace{0mm}
\includegraphics[width=1.0\textwidth]{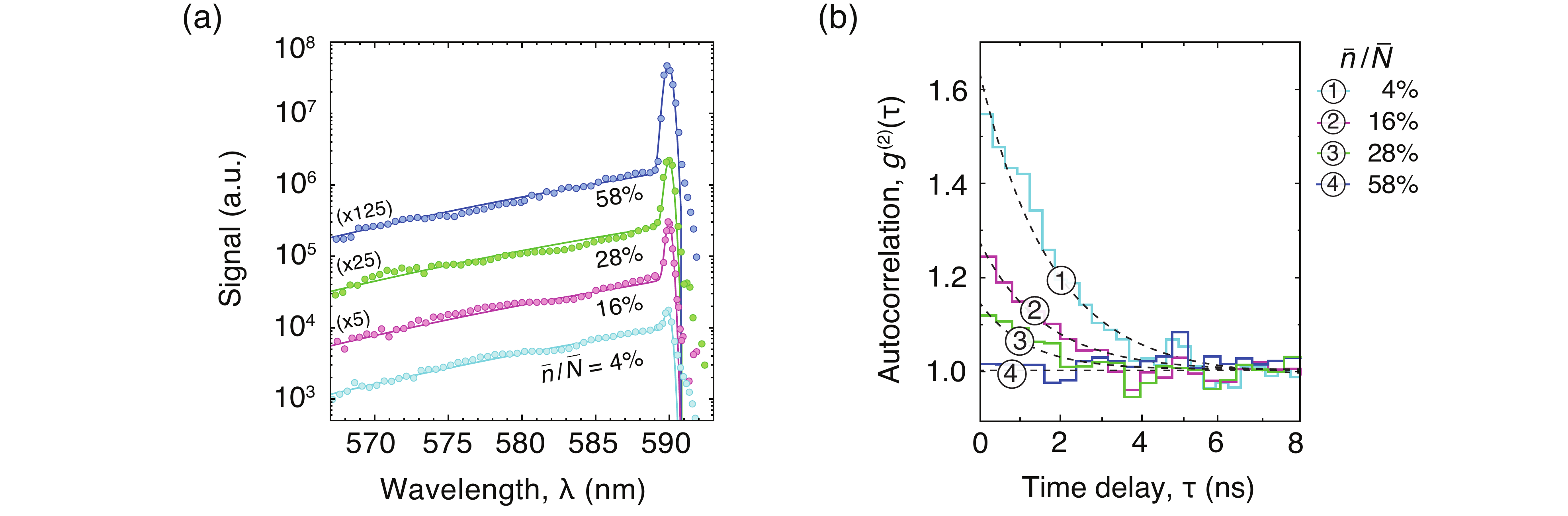}
\par\end{centering}
\vspace{0mm}
\caption{(a) Spectral photon distribution for different condensate fractions (circles), each following a $300~\textrm{K}$ Bose-Einstein distribution (solid lines). Curves have been vertically shifted for clarity. (b) Corresponding second-order correlation functions $g^{(2)}(\tau)$ show photon bunching up to high condensate fractions. Experimental parameters: condensate wavelength $\lambda_\textrm{c} = 2\pi c/\omega_\textrm{c} = 590~\textrm{nm}$ ($\hbar\Delta = -6.7~k_\textrm{\tiny B} T$), dye concentration $\rho = 10^{-3}~\textrm{mol/l}$ (rhodamine 6G).}
\vspace{0mm}
\end{figure}

Typical measurement results for a fixed size of the molecular reservoir are presented in Fig.~4. The thermodynamic state of the photon gas is determined by recording spectra as shown in Fig.~4(a) using a fraction of the light emitted from the cavity. The shown spectra all are in the condensed phase, with an average photon number beyond the critical particle number of $N_\textrm{c}=85\thinspace000$, and show a peak at the wavelength of the cavity cutoff along with a thermal cloud at lower wavelengths. The condensate fraction is obtained by a fit to a $300~\textrm{K}$ Bose-Einstein distribution. Results for the second-order correlation function $g^{(2)}(\tau)$ are shown in Fig.~4(b). While an immediately second-order coherent correlation signal, with $g^{(2)}(0)=1$, would be expected above the BEC transition ($N\geq N_\textrm c$) in case of a strictly conserved particle number, we observe photon bunching, with $g^{(2)}(0)>1$, to extend clearly into the condensed phase regime. For large delays, the observed bunching of the condensate light decays. The expected crossover between the fluctuating regime and the regime with second-order coherence is at
\begin{equation}
\bar n^2 \simeq \frac{M}{\left( 1 + e^{\hbar\Delta/k_\textrm{\tiny B} T} \right)\left( 1 + e^{-\hbar\Delta/k_\textrm{\tiny B} T} \right)} = M_\textrm{eff},
\label{eq1}
\end{equation}
where $M$ denotes the density of dye molecules, $\bar n$ the average number of photons in the condensate mode and $\Delta = \omega_\textrm{c} - \omega_\textrm{zpl}$ the frequency detuning of the condensate from the position of the zero-phonon line of the dye~\cite{Klaers2,Schmitt2}.\\

\begin{figure}[t]
\begin{centering}
\vspace{0mm}
\includegraphics[width=1.0\textwidth]{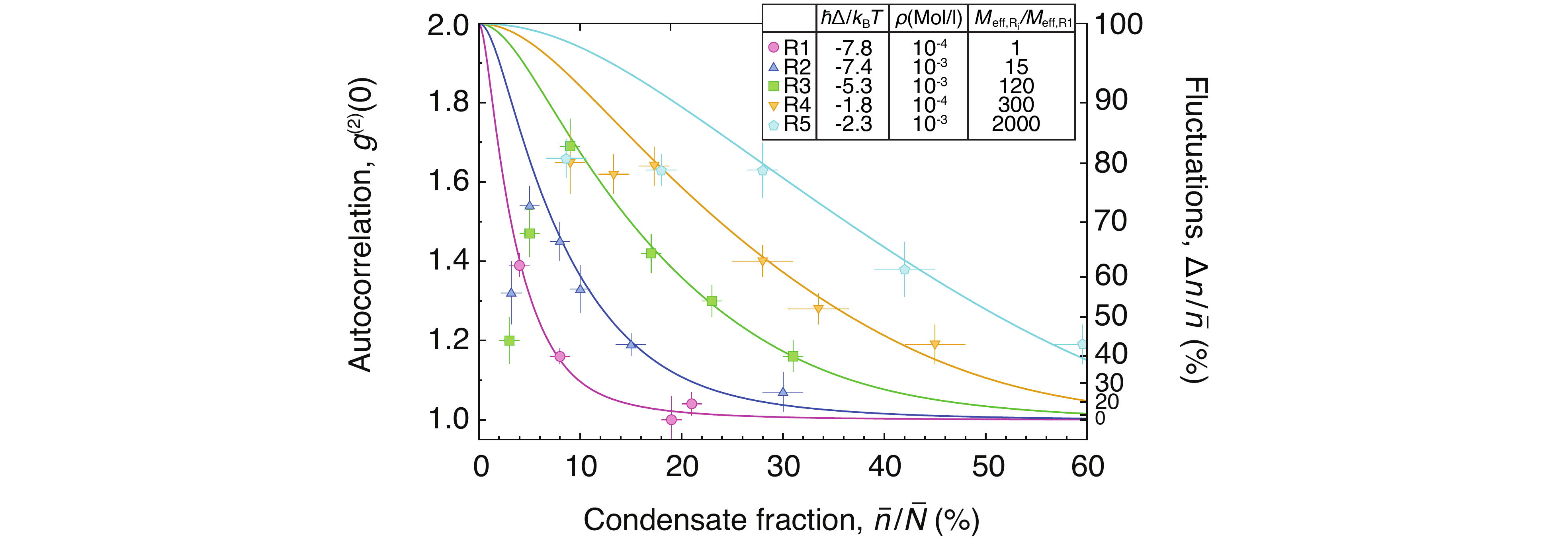}
\par\end{centering}
\vspace{0mm}
\caption{Zero-delay autocorrelations $g^{(2)}(0)$ versus condensate fraction $\bar n/\bar N$ for five different reservoirs of relative effective size $M_{\textrm{eff,R}_i} / M_\textrm{eff,R1}$ with respect to the smallest reservoir R1. Condensate fluctuations extend deep into the condensed phase for high dye concentration $\rho$ and small dye-cavity detuning $\Delta$. Results of a theoretical modeling are shown as solid lines. Experimental parameters: condensate wavelength $\lambda_\textrm{c} = 2\pi c/\omega_\textrm{c} =\{598, 595, 580, 598, 602\}\textrm{nm}$ for data sets R1-R5; dye concentration $\rho =\{10^{-4}, 10^{-3}, 10^{-3}\} \textrm{mol/l}$ for R1-R3 (rhodamine 6G) and $\rho=\{10^{-4}, 10^{-3}\} \textrm{mol/l}$ in R4 and R5 (perylene red).}
\vspace{0mm}
\end{figure}

Figure~5 shows the dependence of the measured zero-delay second-order coherence function $g^{(2)}(0)$ on the condensate fraction for five different combinations of dye concentration and dye-cavity detuning $\Delta$, which varies the effective reservoir size $M_\textrm{eff}$ defined in eq.~(\ref{eq1}). The data sets labelled with R1-R3 have been obtained with rhodamine 6G dye (zero-phonon line at $\omega_\textrm{zpl}\simeq 2\pi c/(545~\textrm{nm})$). For measurements R4 and R5, we have used perylene red dye ($\omega_\textrm{zpl}\simeq 2\pi c/(585~\textrm{nm})$), which allows us to reduce the detuning between condensate and dye reservoir. For the lowest dye concentration and largest detuning (R1), the particle reservoir is so small that the condensate fluctuations are damped soon above the onset of Bose-Einstein condensation (i.e. when the ground mode population becomes macroscopic). The observed emergence of second-order coherence here is attributed to canonical statistical ensemble conditions present in the system. However, by increasing the dye concentration and decreasing the dye-cavity detuning one can systematically extend the regime of large fluctuations to higher condensate fractions. For the largest reservoir realized (R5), where the effective reservoir is approximately 2000 times larger than for reservoir R1, we observe zero-delay correlations of $g^{(2)}(0) \simeq 1.2$ at a condensate fraction of $\bar n/\bar N\simeq 0.6$. The condensate here still performs large relative fluctuations, with $\Delta n/\bar n = \left(g^{(2)}(0)-1\right)^{1/2}\simeq 0.45$, although its occupation number is already comparable to the total photon number. We attribute this as clear evidence that the photon statistics in this system is determined by grand-canonical particle exchange between condensate and dye reservoir. The data agree well with predictions based on a theoretical model shown by solid lines, except when the condensate fraction becomes very small (below some 5\%). In the latter case the measured correlation function reduces towards smaller values, which is attributed to residual light from thermal cavity modes reaching the detection system because of imperfect mode filtering. The averaging over the large number of modes reduces the observed value for $g^{(2)}(0)$ towards unity.\\

\section{Conclusions}

We have described recent experiments with a Bose-Einstein condensate of photons realized in a dye-filled microcavity. The degree of thermalization of the photon gas was varied by tuning the cavity photons closer to resonance or far from resonance with the dye molecules, which in the latter case results in a suppression of thermal contact to the dye and the photons leaking out of the cavity due to mirror losses before they can thermalize, which is reminiscent to usual laser operation. On the other hand, for an enhanced thermal contact with the dye by absorption re-emission processes, photons thermalize to low energetic states near the cavity cutoff and form a Bose-Einstein condensate.\\

In a further series of experiments, the intensity correlations of the photon condensate generated in the dye microcavity system were determined. Relevant to those measurements is that the photo-excitable dye molecules do not only act as a heat bath, but also as an effective particle reservoir for the photon gas. When the dye reservoir is sufficiently large with respect to the system size, we observe grand-canonical statistical fluctuations in the condensed state, while the fluctuations reduce to the usual Poissonian case for a smaller relative size of the reservoir. The results give evidence for Bose-Einstein condensation in the grand-canonical statistical regime.\\

An investigation of the first-order coherence properties of the condensate for different statistical ensemble regimes is subject to current experimental studies. For the future it will be important to test for superfluidity of the photon condensate. An intriguing perspective is the investigation of quantum many-body states in photonic lattices, in which cooling alone can allow for the preparation of entangled many-body states in a thermal equilibrium process when the many-body state is the system ground state.\\\vspace{5mm}

\vspace{-9mm}
\section*{Acknowledgments}
We thank M. Fleischhauer for discussions concerning the laser-BEC comparison of Fig.~2, and acknowledge funding from the ERC (INPEC) and the DFG (We 1748-17).

\bibliographystyle{ws-procs975x65}
\bibliography{ws-pro-sample}

\end{document}